\documentclass[12pt]{article}
\newcommand{\be}{\begin{equation}}
\newcommand{\ee}{\end{equation}}
\newcommand{\ba}{\begin{eqnarray}}
\newcommand{\ea}{\end{eqnarray}}
\newcommand{\se}{\setcounter{equation}{0}}

\newcommand{\1}{^{-1}}
\newcommand{\dg}{^{\dagger}}
\newcommand{\di}{\mbox{d}} 
\newcommand{\e}{\mbox{e}}
\newcommand{\ga}{\gamma_5}

\newcommand{\h}{\frac{1}{2}}
\newcommand{\Id}{\mbox{1\hspace{-1.02mm}l}}   
\newcommand{\la}{\lambda}
\newcommand{\mb}[1]{\quad\mbox{ #1 }\quad}
\newcommand{\mi}[1]{_{\mbox{\tiny #1}}}
\newcommand{\mii}[2]{_{\mbox{\tiny #1}\,#2}}
\newcommand{\mo}[1]{^{(\mbox{\tiny #1})}}
\newcommand{\N}{\tilde{N}}
\newcommand{\Nh}{\hat{N}}
\newcommand{\ra}{\rightarrow}
\newcommand{\re}[1]{(\ref{#1})}
\newcommand{\ve}{\varepsilon} 
\newcommand{\vp}{\varphi}

\begin{document}
\renewcommand{\baselinestretch}{1.05} \small\normalsize

\hfill {\sc HU-EP}-02/03

\vspace*{1.2cm}

\begin{center}

{\Large \bf Form and index of Ginsparg-Wilson fermions} 

\vspace*{0.9cm}

{\bf Werner Kerler}

\vspace*{0.3cm}

{\sl Institut f\"ur Physik, Humboldt-Universit\"at, D-10115 Berlin, 
Germany}
\hspace{3.6mm}

\end{center}

\vspace*{1.2cm}

\begin{abstract}
We clarify the questions rised by a recent example of a lattice Dirac 
operator found by Chiu. We show that this operator belongs to a class based
on the Cayley transformation and that this class on the finite lattice 
generally does not admit a nonvanishing index, while in the continuum limit, 
due to operator properties in Hilbert space, this defect is no longer there. 
Analogous observations are made for the chiral anomaly. We also elaborate on 
various aspects of the underlying sum rule for the index. 
\end{abstract}

\vspace*{0.8cm}

\section{Introduction}
 
\hspace{3mm}    
Recently Chiu \cite{ch01} reported about a lattice Dirac operator, obeying
the Ginsparg-Wilson (GW) relation \cite{gi82} and a mild condition on gauge
field configurations, which does not admit a nonvanishing index, while
everything else, in particular also the chiral anomaly, appears correct. 
The derivation of this was based on the sum rule \cite{ch98} for the index. 
This remarkable observation rises generally the questions of what might be 
missing in the concept and how the apparent paradox resolves. To clarify 
these issues is the aim of the present paper.

In Section 2, inspecting possible forms of unitary operators $V$ which define
GW Dirac operators, we note that the one implicit in Ref.~\cite{ch01} is a 
Cayley transform. In Section 3 we consider the general sum rule for the index
and derive various relations which we need in the following. In Section 4 we 
establish the validity of the sum rule also in Hilbert space. In Section 5 we 
point out the differences to the Atiyah-Singer framework. In Sections 6 we 
generally analyze Cayley-type forms of $V$ on the finite lattice as well as 
in the limit and completely clarify the issues in question. Section 7 contains 
conclusions and summary.

\section{Forms of GW fermions}\se

\hspace{3mm}    
For lattice Dirac operators $D$ which satisfy the GW relation
\be
\{\ga,D\}=\rho\1 D\ga D 
\label{GW}
\ee
and are $\ga$-hermitian
\be
D\dg=\ga D\ga 
\label{ga5}
\ee
due to \re{ga5} the remaining content of \re{GW} is 
$\rho(D+D\dg)=D\dg D=DD\dg$. The latter relation is the condition for unitarity
of the operator $V=\Id-\rho\1 D$, which is also $\ga$-hermitian 
because $D$ is.  
Therefore instead of using \re{GW} and \re{ga5} we equivalently require 
$D$ to have the general form
\be
D=\rho(\Id-V) \mb{ with } V\dg=V\1=\ga V \ga
\label{DN}
\ee
and a real constant $\rho$. Since $V$ is unitary, the spectrum of $D$ is on a 
circle through zero around $\rho$. Real eigenvalues then are possible at $0$ 
and at $2\rho$. Clearly $D$ is normal. 

There are three standard constructions of unitary operators $V$. They are up 
to constant phase factors given
\renewcommand{\labelenumi}{\roman{enumi})}
\begin{enumerate}
\item \label{i} by $X(\sqrt{X\dg X}\,)\1$, normalizing an operator $X$, not
requiring particular mathematical properties of $X$ (apart from $X\dg X\ne0$), 
\item \label{ii} by the Cayley transform $(Y-i\Id)(Y+i\Id)\1$ of
a hermitian operator $Y$, 
\item \label{iii} by $\exp(iZ)$ with a hermitian generator $Z$. 
\end{enumerate}
In addition to unitarity one gets $\ga$-hermiticity of the constructed 
operators $V$ by requiring $X$, $iY$, $iZ$ to be $\ga$-hermitian.

Construction i) is used by Neuberger in the overlap Dirac operator 
\cite{ne98} choosing 
\be
V=-X(\sqrt{X\dg X}\,)\1 \mb{with} X = m\Id + D_W\;,\quad  -2<m<0\;,
\label{DW}
\ee
where $D_W$ is the massless Wilson-Dirac operator and where only $X$ with 
$X\dg X\ne0$ are admitted. This form is well known to give the correct 
results and does not need to be discussed further here. 

With respect to Construction ii) we observe that it is actually the basis of
what has been introduced by Chiu \cite{ch01,ch99}. In fact, the form 
$D=a\1 D_c(\Id+rD_c)\1$ in Ref.~\cite{ch01}, with an appropriate antihermitian 
operator $D_c$ and a suitable positive constant $r\,$, amounts to putting
\be
V=-(Y-i\Id)(Y+i\Id)\1 \mb{ with } Y=irD_c \;,
\label{CH}
\ee
where $\rho\1=2ar$ relates the notations here and there. After a number of 
necessary preparations, we shall analyze Construction ii) in Section 6 in a 
general way and clarify the related questions.

Construction iii) so far has not been applied in the present context. For this
one could use an operator $Z$ of the form of $Y$ mentioned above and tune
$r$ to cover the spectrum appropriately. There remain, however, problems with
the periodicity to be settled. We will not address this case here.

\section{Index and spectrum}\se
 
\hspace{3mm}
To obtain the relations needed on the finite lattice, where $D$ operates in a 
unitary space of finite dimension, we have the spectral representation 
$D=\sum_k\la_kP_k$ of normal $D$ with eigenvalues $\la_k$ and orthogonal 
projections $P_k=P_k\dg\,$. Using the $\ga$-hermiticity of $D$, it gets the 
more detailed form
\be
D=\sum_{j\;(\mbox{\scriptsize Im}\la_j=0)}\la_j (P_j^{(+)}+P_j^{(-)})+
\sum_{j\;(\mbox{\scriptsize Im}\la_j>0)}(\la_j P_j\mo{I}+\la_j^* P_j\mo{II})\;,
\label{specd}
\ee
and one has $P_1^{(+)}+P_1^{(-)}$ projecting on the eigenspace with $\la_1=0$.
Since the projections satisfy $\ga P_j^{(\pm)}=P_j^{(\pm)}\ga= \pm P_j^{(\pm)}$ 
and $\ga P_j\mo{I}=P_j\mo{II}\ga$, one gets
\be
\mbox{Tr}(\ga P_j^{(\pm)})=\pm N_{\pm}(\la_j) \;,\quad
\mbox{Tr}(\ga P_j\mo{I})=\mbox{Tr}(\ga P_j\mo{II})=0\;. 
\label{TP}
\ee
where the trace Tr is in full space and $N_{+}(\la_j)$ and $N_{-}(\la_j)$ 
denote the dimensions of the right-handed and of the left-handed eigenspace 
for real eigenvalue $\la_j$, respectively. 

Using \re{specd} with \re{TP} the index $N_+(0)-N_-(0)$ is 
obtained from the resolvent $(D-\zeta\,\Id)\1$ by
\be 
\lim_{\zeta\rightarrow 0}\mbox{Tr}\Big(\ga(-\zeta)(D-\zeta\,\Id)\1 \Big)= 
N_+(0)-N_-(0)
\label{IND}
\ee
and one also finds
\be 
\lim_{\zeta\rightarrow 0}\mbox{Tr}\Big(\ga D(D-\zeta\,\Id)\1 \Big)  
=\sum_{\la_j\ne0\mbox{ \scriptsize real }}\Big(N_+(\la_j)-N_-(\la_j)\Big)\;.
\label{IND1}
\ee
Adding up Eqs.~\re{IND} and \re{IND1} the sum of their l.h.s.~gets 
$\mbox{Tr}(\ga\Id)=0$ and one obtains
\be  
0=N_+(0)-N_-(0)
+\sum_{\la_j\ne0\mbox{ \scriptsize real }}\Big(N_+(\la_j)-N_-(\la_j)\Big)\;.
\label{sum}
\ee
This shows that to make a nonvanishing index at all possible, one has to 
require that the spectrum of $D$ in addition to zero allows for at least one 
further real eigenvalue.  

The sum rule found by Chiu in Ref.~\cite{ch98} and used in Ref.~\cite{ch01}
is the special case of \re{sum},
\be
N_+(0)-N_-(0)+ N_+(2\rho)-N_-(2\rho)=0\;,
\label{sums}
\ee
which holds for Dirac operators satisfying the GW relation \re{GW}. In that 
case \re{IND1} simplifies to 
\be
(2\rho)\1\mbox{Tr}(\ga D)=N_+(2\rho)-N_-(2\rho)\;.
\label{IND1s}
\ee

In the following it will be more appropriate to have the relations in terms 
of $V$, in which case $\rho$ drops out completely. For this we note that by 
\re{DN} the eigenvalues $0$ and $2\rho$ of $D$ correspond to the eigenvalues 
$1$ and $-1$ of $V$, respectively. Thus, denoting the dimensions of the 
right-handed and the left-handed eigenspace at eigenvalue $\pm1$ of $V$ by 
$\N_{+}(\pm1)$ and $\N_{-}(\pm1)$, respectively, the sum rule \re{sums}  
turns into 
\be
\N_+(1)-\N_-(1)+\N_+(-1)-\N_-(-1)=0\;,
\label{vS}
\ee
requiring that in addition to $+1$ the eigenvalue $-1$ of $V$ can occur.
Further, instead of \re{IND1s} with \re{sums}, we then have for the index 
\be
\N_+(1)-\N_-(1)=-\N_+(-1)+\N_-(-1)=\h\mbox{Tr}(\ga V)\;.
\label{IND1v}
\ee

In this context it is instructive to note that the spectral representation
of $\ga V$ is
\be
\ga V= P_1^{(+)}-P_1^{(-)}-P_2^{(+)}+P_2^{(-)}+
              \sum_k(\check{P}_k^{[+]}-\check{P}_k^{[-]})
\label{dV}
\ee
where $P_1^{(\pm)}$ and $P_2^{(\pm)}$ project onto the eigenspaces of $V$ with 
eigenvalue $+1$ and $-1$, respectively, and
where $\check{P}_k^{[\pm]}=\h(P_k\mo{I}+P_k\mo{II}\pm\e^{i\vp_k}\ga P_k\mo{I}
\pm\e^{-i\vp_k}\ga P_k\mo{II})$ with $0<\vp_k<\pi\,$ (and $[\pm]$ being related
to signs of eigenvalues of $\ga V$).

\section{Sum rule in Hilbert space}\se
 
\hspace{3mm}
For the clarification of the questions raised in Ref.~\cite{ch01} we have to 
make sure that the sum rule still holds in the continuum limit. Then the 
operators act in Hilbert space (as will be discussed in detail in Section 6),
so that we firstly have to give a definition of the traces Tr in infinite 
space, too, and secondly to show that a continuous part of the spectrum does 
not contribute to the sum rule.

The immediate prescription for the Tr-expressions is to consider them
as the limit of the respective finite-lattice results. Technically more 
convenient is summing or integrating up a spectral representation {\it after} 
taking Tr of its contributions. Clearly any regularization can be used by which
one arrives at the limit of the finite-lattice results.

The Tr-expressions here typically involve differences of operators projecting 
on right-handed and left-handed space, respectively. This suggests to use the
notion of cardinalities, which extends the one of dimensions to the infinite 
case. It allows to compare infinite spaces directly, using that two sets have 
the same cardinality if there is a bijective mapping between them. For our 
purpose we restrict such mappings to ones between subspaces which have 
different chirality and call the resulting concept specialized cardinality. 

A simple example is $\mbox{Tr}(\ga\Id)=0$, which we have used to derive 
\re{sum}. In the infinite case one can decompose the identity operator as 
$\Id=\sum_j \hat{P}_j\,$, with any orthogonal projections $\hat{P}_j$ having 
finite trace Tr and being trivial in Dirac space, and define $\mbox{Tr}(\ga\Id)
=:\sum_j \mbox{Tr}(\ga \hat{P}_j) =0$. The special choice $\hat{P}_n=\Id_0
\otimes p(n)$, where $p(n)$ is the projection related to lattice site $n$ and 
$\Id_0$ the identity operator in Dirac and gauge field space, allows to make 
contact to the finite-lattice results. With respect to the general meaning,
for $\ga\Id=P\mi{R}-P\mi{L}$ with the projections $P\mi{R}=\h(1+\ga)\Id$ and 
$P\mi{L}=\h(1-\ga)\Id$ a bijective mapping is provided by $T= \gamma_4\Id$, 
which gives $T\dg P\mi{R} T=P\mi{L}$ and $T\dg P\mi{L} T=P\mi{R}$. This 
indicates that $P\mi{R}$ and $P\mi{L}$ project on subspaces which have the 
same specialized cardinality.

A further example is $\h\mbox{Tr}(\ga V)$ in \re{IND1v}. The discrete part of
the spectral representation of the operator $\ga V$ is given by \re{dV}, while 
the continuous part does not contribute (as will be shown below). One notes 
that the subspaces on which $\sum_k \check{P}_k^{[+]}$ and 
$\sum_k\check{P}_k^{[-]}$ project have the same specialized cardinality. This 
follows because the projections $\check{P}_k^{[+]}$ and $\check{P}_k^{[-]}$ 
are one-to-one associated and the subspaces on which such pairs project have 
both dimension $\h\mbox{Tr}(P_k\mo{I}+P_k\mo{II})\,$. Thus only the
part with $P_1^{(\pm)}$ and $P_2^{(\pm)}$ contributes. 

We next address the fact that the spectrum of $D$ now can also have a 
continuous part. The discrete part of the spectral representation is as before.
The continuous part in terms of $V$ is represented by a Stieltjes integral,
\be
V\mi{con}=\int_{-\pi}^{\pi}\e^{i\vp}\,\di E\mii{con}{\vp}
         =\int_{0}^{\pi}\e^{i\vp}\,\di E\mii{con}{\vp}
         -\int_{0}^{\pi}\e^{-i\vp}\,\di E\mii{con}{-\vp}\;,
\label{Dcon}
\ee
where the subdivision into two integrals is possible because the projector 
function $E\mii{con}{\vp}$ is purely continuous. With $V\dg=\ga V\ga$ and 
$E\mii{con}{\vp}\dg=E\mii{con}{\vp}$ one gets $\ga E\mii{con}{\vp}=
-E\mii{con}{-\vp}\ga\,$. From this, using that $E\mii{con}{\vp'}E\mii{con}{\vp}
=E\mii{con}{\vp}E\mii{con}{\vp'}=E\mii{con}{\vp}$ holds for $\vp\le\vp'$, it 
follows that $\mbox{Tr}(\ga  E\mii{con}{\vp})= 0\,$. 
Therefore we obtain for any function 
$F(V\mi{con})$ 
\be
\mbox{Tr}\Big(\ga F(V\mi{con})\Big)=\int_{-\pi}^{\pi}F(\e^{i\vp})\,\di 
\Big(\mbox{Tr}(\ga E\mii{con}{\vp})\Big)=0
\label{TT}
\ee
and an analogous relation for the continuous part of $D$. Because the sum rule
is based on expressions of type $\mbox{Tr}\Big(\ga F(V)\Big)$ or 
$\mbox{Tr}\Big(\ga F(D)\Big)$, it is indeed {\it not} affected by the 
occurrence of a continuous spectrum.

\section{Differences to the Atiyah-Singer case}\se
 
\hspace{3mm}
The sum rule, which in Ref.~\cite{ch01} and in the present paper is seen to be
of crucial importance, exhibits the mechanism allowing for a nonvanishing
index on the lattice. This mechanism is related to the chiral noninvariance, 
which is conceptually necessary in quantum field theory, and in its lattice 
definition is present from the start. Clearly in lattice theory the space 
structure itself neither does depend on the Dirac operator nor does it get 
chirally asymmetric. 

This is in contrast to what one has in case of the Atiyah-Singer Dirac operator 
\cite{at68}, where to admit a nonvanishing index the space structure itself
must be allowed to depend on the particular gauge field configuration. Because 
this appears to be not sufficiently realized, we briefly point out some 
details. 

The definition in the Atiyah-Singer case is based on (Weyl)
operators $D^{(+)}\mi{AS}$ and $D^{(-)}\mi{AS}$ which map from the total 
right-handed space ${\cal E}^{(+)}$ to the total left-handed space 
${\cal E^{(-)}}$ and back, respectively. It is given in the combined space 
${\cal E}^{(+)}\oplus {\cal E^{(-)}}$ by $D\mi{AS}=\hat{D}^{(+)}\mi{AS}+
 \hat{D}^{(-)}\mi{AS}$ where $\hat{D}^{(\pm)}\mi{AS}=D^{(\pm)}\mi{AS}$ on 
${\cal E^{(\pm)}}$ and $\hat{D}^{(\pm)} \mi{AS}=0$ on ${\cal E^{(\mp)}}$. 
Because of $D^{(+)\dag}\mi{AS}=D^{(-)}\mi{AS}$ the Dirac operator $D\mi{AS}$ 
is self-adjoint and since it acts on a compact manifold its spectrum is 
discrete. Thus it is represented by $D\mi{AS}=\sum_j\la\mii{AS}{j}
P\mii{AS}{j}$ which implies $D\mi{AS}^2=\sum_j\la\mii{AS}{j}^2P\mii{AS}{j}$. 

On the other hand, one gets $D\mi{AS}^2=\hat{D}^{(+)}\mi{AS}\hat{D}^{(-)}
\mi{AS}+\hat{D}^{(-)}\mi{AS}\hat{D}^{(+)}\mi{AS}$, which can be evaluated
noting that the operators $D^{(-)}\mi{AS}D^{(+)}\mi{AS}$ and $D^{(+)}\mi{AS}
D^{(-)}\mi{AS}$ map within ${\cal E}^{(+)}$ and ${\cal E}^{(-)}$, respectively,
and are selfadjoint and nonnegative. With the eigenequation $D^{(-)}D^{(+)}
\Phi_j=\kappa_j\Phi_j$ in ${\cal E}^{(+)}$ one gets the eigenequation 
$D^{(+)}D^{(-)}(D^{(+)}\Phi_j)=\kappa_j(D^{(+)}\Phi_j)$ in ${\cal E}^{(-)}$. 
Further, from $\langle D^{(+)}\Phi_{j r'}|D^{(+)}\Phi_{j r}\rangle=
\langle\Phi_{j r'}|D^{(-)}D^{(+)}\Phi_{j r}\rangle=\kappa_j\langle\Phi_{j r'}|
\Phi_{j r}\rangle$ one sees that, except for $\kappa_j=0$, for each of the 
common eigenvalues $\kappa_j$ the eigenspaces must have the same dimension. 
Therefore, except for $\kappa_j=0$, the operators $D^{(-)}\mi{AS}D^{(+)}
\mi{AS}$ and $D^{(+)}\mi{AS}D^{(-)}\mi{AS}$ have the same spectra. Denoting 
the projections on their eigenspaces by $P\mii{AS}{j}^{(+)}$ and 
$P\mii{AS}{j}^{(-)}$, respectively, and comparing the above expressions for 
$D\mi{AS}^2$ it then follows that 
$D\mi{AS}=\sum_j\la\mii{AS}{j}(P\mii{AS}{j}^{(+)}+P\mii{AS}{j}^{(-)})$ 
and that one always has 
\be
\Nh_{+}(\la\mii{AS}{j})=\Nh_{-}(\la\mii{AS}{j})\mb{ for }\la\mii{AS}{j}\ne0
\label{as1}
\ee
for the dimensions $\Nh_{+}(\la\mii{AS}{j})=\mbox{Tr}\,P\mii{AS}{j}^{(+)}$ and
$\Nh_{-}(\la\mii{AS}{j})=\mbox{Tr}\,P\mii{AS}{j}^{(-)}$ of the eigenspaces. 

For $\kappa_j=\la\mii{AS}{j}=0$ one gets
ker$\,D^{(+)}\mi{AS}=$ ker$\,D^{(-)}\mi{AS}D^{(+)}\mi{AS}$ and 
ker$\,D^{(-)}\mi{AS}=$ ker$\,D^{(+)}\mi{AS}D^{(-)}\mi{AS}$ (as is obvious 
from left to right and follows from right to left from 
$\langle\Phi|D^{(-)}\mi{AS}D^{(+)}\mi{AS}\Phi\rangle=
\langle D^{(+)}\mi{AS}\Phi|D^{(+)}\mi{AS}\Phi\rangle$). 
Thus the eigenspaces with eigenvalue zero have the dimensions
$\Nh_{\pm}(0)=$ dim~ker$\,D^{(\pm)}\mi{AS}$ and the index becomes $\Nh_+(0)-
\Nh_-(0)=\mbox{dim ker }D^{(+)}\mi{AS}-\mbox{dim ker }D^{(+)\dag}\mi{AS}$. 

It is now seen that because of \re{as1}, to admit a nonvanishing index 
$\Nh_{+}(0)-\Nh_{-}(0)$, the space structure itself must be allowed to vary 
and to get chirally asymmetric, and thus to depend on the particular gauge 
field configuration. It also becomes obvious that in case of a nonvanishing
index the specialized cardinalities of ${\cal E}^{(+)}$ and ${\cal E}^{(-)}$ 
get different. Indeed, while right-handed and left-handed eigenspaces with the
same dimension are associated at each $\la\mii{AS}{j}\ne0$, this then does not
hold for those with $\la\mii{AS}{j}=0$. 

In addition to the fundamentally different space structure, a further 
difference is that in the Atiyah-Singer framework one considers a 
differential operator on a compact manifold, while in the definition of 
quantized theory one deals with a lattice operator and its subtle limit on 
{\bf R}$^4$. Thus altogether special care appears appropriate when considering
 analogies.

\section{Analysis of Cayley-type $V$}\se

\hspace{3mm}    
To investigate Construction ii) of Section 2 we consider the general choice
\be
V=-(Y-i\Id)(Y+i\Id)\1=2(\Id+Y^2)\1-\Id+i\,2Y(\Id+Y^2)\1\;.
\label{CA}
\ee
In \re{CA} for the eigenvalue $y=0$ of $Y$ one obviously gets the eigenvalue 
$v=1$ of $V$. 

On the finite lattice the operators act in a unitary space of finite dimension.
Therefore, requiring $Y$ to be a well-defined hermitian operator, its spectrum
consists of a finite number of real eigenvalues (which are discrete and 
finite). Introducing $s=\mbox{max}(|y\mi{min}|,|y\mi{max}|)$, where $y\mi{min}$
and $y\mi{max}$ denote the smallest and the largest eigenvalue of $Y$, 
respectively, \re{CA} gives
\be
\mbox{Re }v\ge\frac{2}{1+s^2}-1\;\mbox{ for all }v\;,\quad
|\mbox{Im }v|\ge \frac{2s}{1+s^2}\;\mbox{ for }\;\mbox{Re }v<0\;.
\label{CC}
\ee
We see from this that the eigenvalue $v=-1$ of $V$ cannot be reached. Thus 
on the finite lattice Construction ii) does generally not meet the basic 
requirement needed according to \re{vS} to allow for a nonvanishing index. 

In addition, we observe that , because of not reaching $v=-1$, one always has 
$P^{(\pm)}_2 \equiv0$ in \re{dV}. Thus on the finite lattice already 
the chiral anomaly turns out to be affected.

The obvious obstacle which prevents from reaching the eigenvalue $-1$ of $V$ 
is that on the finite lattice $Y$ is bounded. A related problem is that there 
the inverse Cayley transform, 
\be
Y=-i(V-\Id)(V+\Id)\1\;,
\label{BA}
\ee 
is not valid for all unitary operators but only for the subset for which the
spectrum does not extend to $-1\,$.

The crucial observation now is that the indicated restrictions no longer hold
in Hilbert space, where one gets a well-defined connection between general 
unitary operators $V$ and selfadjoint operators $Y$ which can also be 
unbounded. To recall how this comes about we start from the general spectral 
representation of unitary operators,
\be
V=\int_{-\pi}^{\pi}\e^{i\vp}\di E_{\vp}\;, 
\label{EV}
\ee 
where the projection function $E_{\vp}$ accounts for discrete as well as for 
continuous contributions. Naive insertion of \re{EV} into \re{BA} does not 
generally make sense because $-i(\e^{i\vp}-1) (\e^{i\vp}+1)\1 =\tan\!
\frac{\vp}{2}$ is not bounded, diverging for $\vp=\pm\pi$ where the value $-1$
of the spectrum of $V$ is reached. However, $Y$ is well-defined on Hilbert 
space vectors $f$ by 
\be
Yf=\lim_{{\tilde\vp}\ra\pi}\int_{-\tilde{\vp}}^{\tilde{\vp}}
\tan \!\frac{\vp}{2} \,\di (E_{\vp}f)
\label{EY}
\ee
in the sense of strong convergence. This is seen noting that with 
$f=(\Id+V)g$ one gets $\tan^2\!\frac{\vp}{2}\,\di||E_{\vp}f||^2=4\sin^2\!
\frac{\vp}{2}\,\di||E_{\vp}g||$ over which the integral from $-\pi$ to $\pi$ 
is obviously finite.

Thus with unbounded operators $Y$ in \re{EY} we indeed get unitary operators
$V$ in \re{EV} with a spectrum extending to $-1\,$, as is necessary for proper
working of the sum rule \re{vS} and to guarantee correct results in \re{IND1v}.
It is seen that for this a Hilbert space is necessary, which not only has 
infinite dimension but also includes its limit elements. Obviously the latter 
here is of crucial importance.

Since the projection function $E_{\vp}$ is common to \re{EV} and \re{EY}, we 
can trace back for $V$ as well as for $Y$ to the projections $P_1^{(\pm)}$ and 
$P_2^{(\pm)}$ (related to the eigenvalues $+1$ and $-1$ of $V$, respectively)
which only enter the sum rule. With $E_{\vp}$ being 
strongly continuous from the right, $\lim_{\ve\ra0}E_{\vp+0}=E_{\vp}\,$, for 
this one has the relations $\lim_{\ve\ra0}(E_{0}-E_{-\ve})=P_1^{(+)}+P_1^{(-)}$
and $\lim_{\ve\ra0}(E_{\pi}-E_{\pi-\ve})=P_2^{(+)}+P_2^{(-)}$. 

To see how the Hilbert space of interest arises in the continuum limit, one has
to reconsider things from the appropriate point of view$\,$: On the 
infinite lattice one gets the Hilbert space of sequences ${\bf l}_2\;$ (in a
basis of which a vector is related to a lattice site, a Dirac index and a gauge 
field index). The unitarily equivalent Hilbert space ${\bf L}_2(\pi;k)$ of 
functions $f(k)$ with $-\pi\le k_{\mu}\le\pi$ (and Dirac and gauge-group 
indices being suppressed) is obtained from ${\bf l}_2$ by a Fourier 
transformation.  Introducing the lattice spacing $a$ and variables $p=k/a$ the
 space ${\bf L}_2(\pi;k)$ becomes ${\bf L}_2(\pi/a;p)$. By the limit $a\ra0$ 
one then gets the operators in ${\bf L}_2(\infty;p)$ from the ones in 
${\bf L}_2(\pi/a;p)$. The space ${\bf L}_2(\infty;x)$, unitarily equivalent to
${\bf L}_2(\infty;p)$, is again obtained  by a Fourier transformation. Instead 
of proceeding in the more instructive way sketched, one can also realize the 
direct way from ${\bf l}_2\;$ to ${\bf L}_2(\infty;x)$. The equivalent spaces 
${\bf L}_2(\infty;x)$ and ${\bf L}_2(\infty;p)$ are the ones one has in the 
continuum limit. 

In detail the definition of the operators of interest by the indicated limit
needs some care. Firstly, since the limit element cannot be given explicitly,
we resort to the definition by all matrix elements, i.e.~by weak operator 
convergence. Secondly, because two spaces are involved, this weak limit is to 
be slightly generalized. To show that this can be properly done, we introduce 
$f_a(p)=f(p)\Pi_{\mu}\Theta(\pi/a-p_{\mu})$ with $f(p)\in{\bf L}_2(\infty;p)$ 
and the operator $\hat{{\cal O}}_a\,$, requiring $\hat{{\cal O}}_a(p',p)$ of 
${\bf L}_2(\infty;p)$ to be equal to ${\cal O}_a(p',p)$ of ${\bf L}_2(\pi/a;p)$ 
for $-\pi/a\le p_{\mu}\le\pi/a$. Then $\langle f_a|\hat{{\cal O}}_a g_a\rangle$
in ${\bf L}_2(\infty;p)$ equals $\langle f_a|{\cal O}_a g_a\rangle$ in 
${\bf L}_2(\pi/a;p)$ for all finite $a$ and $\langle f_a|\hat{{\cal O}}_a 
g_a\rangle\ra\langle f|\hat{{\cal O}}g\rangle$ for $a\ra0$ defines the operator
$\hat{{\cal O}}$ in ${\bf L}_2(\infty;p)$.
 
Thus, though we cannot give explicit expressions for the operators in 
${\bf L}_2(\infty;p)$ and ${\bf L}_2(\infty;x)$, we can unambiguously deal with
them$\,$: First calculating the desired matrix elements of the operators or of 
the functions of operators of interest on the infinite lattice and then 
performing the $a\ra0$ limit leads to the results in terms of the respective 
matrix elements. It is seen here that what is practice in correct approaches
can be precisely formulated and justified in Hilbert space.

The mapping from ${\cal O}_a$ to $\hat{{\cal O}}$ is not invertible and the 
spectra of ${\cal O}_a$ and $\hat{{\cal O}}$ can be substantially different. 
Obviously the momenta get unbounded in the limit, and this also holds for the 
operators $Y$ of physical interest. Therefore by the connection between \re{EY}
and \re{EV} the spectrum of $V$ can extend to $-1\,$, as has been shown to be 
necessary for correct results.

We now see how the paradox of Ref.~\cite{ch01} resolves$\,$: The chiral anomaly
there, being already a continuum quantity, should not be considered with the 
sum rule on the finite lattice but with the one valid in the continuum limit. 
In this limit the chiral anomaly as well as the sum rule can get correct
properties, while on the finite lattice both of them are affected by not 
reaching the eigenvalue $-1$ of $V$. The problems on the finite lattice have 
been seen to occur not only in the example of Ref.~\cite{ch01} but for any 
well-defined $Y$.

\section{Conclusions and summary}\se
 
\hspace{3mm}
Inspecting possible forms of unitary operators $V$ which define GW Dirac 
operators, we have pointed out that the Dirac operator proposed by Chiu is 
based on an operator $V$ of the Cayley-transform type. After a number of
necessary preparations, we have analyzed the properties of this class of
operators. 

Considering the general sum rule for the index we have derived various 
relations in finite space which have been important in the following. To make
sure that the sum rule still holds in the continuum limit, we firstly have 
shown that the extension to infinite space does not spoil it and secondly that
a continuous part of the spectrum does not contribute to it. 

We have stressed that the mechanism for admitting a nonvanishing index, 
exhibited by the sum rule here, is different from that in the Atiyah-Singer 
case, where instead the space structure itself is allowed to depend on the 
particular gauge field configuration.

Considering Cayley-type operators $V$ we have seen that for them on the finite
lattice the sum rule generally does not allow for a nonvanishing index (and not 
only in the example of Chiu) because the eigenvalue $-1$ of $V$ cannot be 
reached. In addition, we have seen that in case of such $V$ on the finite
lattice already the chiral anomaly is affected by this.

Turning to Hilbert space we have pointed out that there the Cayley 
transformation provides a well-defined connection between unbounded self-adjoint
operators and general unitary operators $V$ for which the eigenvalue $-1$ can 
be reached. We then have shown that, after introducing a slightly generalized 
weak operator convergence, the continuum limit precisely provides the 
appropriate Hilbert-space formulation. 

Our results make clear that in the continuum limit the problems
of the sum rule for the index and of the chiral anomaly with Cayley-type 
operators $V$ disappear. The paradox of Ref.~\cite{ch01} has been seen to 
resolve in that the chiral anomaly there, being already a continuum quantity, 
should not be considered with the sum rule on the finite lattice but with the 
one valid in the limit.

\section*{Acknowledgements}

\hspace{3mm}
I would like to thank Ting-Wai Chiu for drawing my attention to these problems.
I am grateful to Michael M\"uller-Preussker and his group for their kind 
hospitality.

\renewcommand{\baselinestretch}{0.8} 

\end{document}